\documentstyle[sprocl]{article}

\def\PLB{{\em Phys. Lett.}  B}

\def\PRD{{\em Phys. Rev.} D}

\def\be{\begin{equation}}
\def\ee{\end{equation}}
\def\bea{\begin{eqnarray}}
\def\eea{\end{eqnarray}}

\begin{document}

\title{Gauge-Invariant QCD --- an Approach to Long-Range Color-Confining Forces.}

\author{Kurt Haller}

\address{Department of Physics, University of Connecticut, Storrs, CT 06269}

\maketitle\abstracts{We discuss the transformation of the QCD temporal-gauge Hamiltonian to a representation in
which it can be expressed as a functional of gauge-invariant quark and gluon fields. We show how this objective 
can be realized by implementing the non-Abelian Gauss's law, and by using the mathematical apparatus developed 
for that purpose to also construct gauge-invariant quark and gluon fields. We demonstrate that,
in the transformed QCD Hamiltonian, the interactions of pure-gauge components of the gauge field with 
color-current densities are replaced by nonlocal interactions connecting 
quark color-charge densities to each other and to `glue'-color.
We discuss the nonperturbative evaluation of these nonlocal interactions, which are non-Abelian analogs of the 
Coulomb interaction in QED, and we explore their implications for QCD in the low-energy regime.}

\section{Introduction}
The problem that my students\footnote{ Lusheng Chen, and Mario Belloni, Ph.D., University of Connecticut, 1997.} and I addressed in the investigation I am discussing in this report, deals
with the role of the non-Abelian Gauss's law in generating a nonlocal 
interaction between gauge-invariant color-bearing particle states --- in particular, quark and 
multi-quark states. Our conjecture that this nonlocal interaction might play an 
important, perhaps a decisive, role in color confinement, is 
based on an analogy with Abelian gauge theories such as 
QED. In covariant-gauge QED, the spinor (electron) field $\psi$ is a gauge-dependent quantity. It
interacts with `pure gauge' components of the gauge field $A^{\mu}$, and that interaction is responsible 
for generating --- perturbatively, order by order --- the parts of the S-matrix elements that correspond to 
the Coulomb interaction between charges. On the other hand, when covariant-gauge QED is transformed to a 
representation in which  Gauss's law 
is implemented and $\psi$ becomes a gauge-invariant charged field, the same nonlocal 
Coulomb interaction that is seen in the Coulomb gauge appears explicitly in that case,
even though the covariant-gauge condition continues to apply.~\cite{khelqed} 
Quite generally, when only gauge-invariant fields are used in constructing the QED Hamiltonian, 
the interactions between charged fields and pure-gauge components of gauge fields vanish, 
and a nonlocal interaction between charge densities --- the Coulomb interaction ---  appears in their stead.
In gauge-invariant QED, 
the only interaction besides the Coulomb interaction is between the transverse (gauge-invariant) 
part of the gauge field and the current density; this interaction takes the form 
$-\,{\int}j_i({\bf r})A_{Ti}\,({\bf r})d{\bf r}$, where $A_{Ti}$ is the transverse (gauge-invariant) 
part of the gauge field, and 
$\vec{j}({\bf r})=e{\psi^\dagger}({\bf r})\vec{\alpha}\psi({\bf r})$. Since the current density has a $v/c$ 
dependence in the nonrelativistic limit, the Coulomb interaction is,
by far, the most important electrodynamic force in the low-energy regime. 
That fact provides a strong incentive for formulating QCD in 
terms of gauge-invariant fields, since then, in QCD too, the interactions between the `pure gauge' components of the 
gauge field and the color-current density must necessarily be eliminated and replaced by a nonlocal interaction between color-charge 
densities. We have obtained such a nonlocal interaction and, in the remainder of this paper, will  
discuss how it comes about and how it can facilitate our study of the low-energy regime of QCD.
\section{Implementing the Non-Abelian Gauss's Law}
The strategy of using only gauge-invariant fields in formulating gauge theories, 
in QCD as well as in QED,  faces major technical hurdles. The most daunting is the implementation 
of the non-Abelian Gauss's law. The Gauss's law operator for QCD is 
\begin{equation}
{\hat {\cal G}}^{a}({\bf r})=\partial_{i}\Pi_{i}^{a}({\bf r})+gf^{abc}A_{i}^{b}({\bf r})\,,
\Pi_{i}^{c}({\bf r})+j_{0}^{a}({\bf r})\,,
\label{eq:QCDGauss}
\end{equation}
where $\Pi_{i}^{a}$ is the negative chromoelectric field and, also, the momentum conjugate to  
$A_{i}^{a}({\bf r})$; $j^a_{0}$ is the color-charge density $j^a_{0}({\bf{r}})=g\,\,\psi^\dagger({\bf{r}})\,
{\textstyle\frac{\lambda^a}{2}}\,\psi({\bf{r}})\,,$ and $f^{abc}$ represents the SU(3) structure constants. 
The Abelian Gauss's law operator for QED is 
\begin{equation}
{\hat {\sf G}}({\bf r})=\partial_{i}\Pi_{i}({\bf r})+j_{0}({\bf r})\,,
\label{eq:QEDGauss}
\end{equation}
where $j_{0}$ is the electric charge density $j_{0}({\bf{r}})=g\,\,\psi^\dagger({\bf{r}})\,
\psi({\bf{r}}).$ In QED, ${\hat {\sf G}}({\bf r})$ and $\partial_{i}\Pi_{i}({\bf r})$ 
are unitarily equivalent, which makes it far easier to implement Gauss's law in QED than in QCD. 
We note, for example, that $\partial_{i}\Pi_{i}({\bf r})\,|{\xi}\rangle=0$, 
or an equivalent equation, is easy to solve. Moreover, it is  
relatively straightforward to explicitly construct a unitary transformation that relates 
${\hat {\sf G}}$ and $\partial_{i}\Pi_{i}({\bf r})$.~\cite{qedimp} That unitary 
transformation enables us to transform 
solutions of $\partial_{i}\Pi_{i}({\bf r})\,|\xi\rangle=0$ into solutions of 
${\hat {\sf G}}({\bf r})\,|{\hat {\xi}}\rangle=0$, thereby implementing Gauss's law. In QCD, however,
${\hat {\cal G}}^{a}$ and $\partial_{i}\Pi_{i}^{a}$
{\em cannot} be unitarily equivalent, because their commutator algebras differ. Whereas 
$\left[\partial_{i}\Pi_{i}^{a}({\bf r}), \partial_{i}\Pi_{i}^{b}({\bf r}^{\prime})\right]=0$, 
$\left[{\hat {\cal G}}^a({\bf{r}}),
{\hat {\cal G}}^b({\bf{r^\prime}})\right]=
igf^{abc}{\hat {\cal G}}^c({\bf{r}){\delta}({\bf r-r^\prime}})\,.$~\cite{khtemp} Although 
$\partial_{i}\Pi_{i}^{a}({\bf r})\,|{\xi}\rangle=0$ is just as easy to solve as its Abelian counterpart, 
we cannot make use of that solution to implement the non-Abelian Gauss's law
\be
{\hat {\cal G}}^{a}({\bf r})\,|{\hat {\Psi}}\rangle=0\,.
\label{eq:impnagauss}
\ee
 That fact requires us to use 
a fundamentally different strategy for implementing Gauss's law in QCD. \smallskip

We have constructed 
states that satisfy Gauss's law in QCD and Yang-Mills theory, by solving an operator-valued differential 
equation.~\cite{BCH1,CBH2} In that work, we defined an operator-valued quantity 
$\overline {{\cal A}_{i}^{\gamma}}({\bf r})$, which we will call the {\em resolvent gauge field}, and  which obeys the
operator differential equation 
\begin{eqnarray}
i\int d{\bf r}^\prime&&[\,\partial_{i}\Pi_{i}^{a}({\bf r}),\,
\overline{{\cal A}_{j}^{\gamma}}({\bf r}^\prime)\,]\,
V_{j}^{\gamma}({\bf r}^\prime)
+ igf^{a\beta d}A_{i}^{\beta}({\bf r})\int d{\bf r}^\prime
[\,\Pi_{i}^{d}({\bf r}),\,
\overline{{\cal A}_{j}^{\gamma}}({\bf r}^\prime)\,]\,
V_{j}^{\gamma}({\bf r}^\prime)
\nonumber \\
&&= -gf^{a\mu d}\,A_{i}^{\mu}({\bf r})\,V_{i}^{d}({\bf r})
\nonumber \\
&&+\sum_{\eta=1}{\textstyle\frac{g^{\eta +1}B(\eta)}{\eta!}}\,
f^{a\beta c}f^{\vec{\alpha}c\gamma}_{(\eta)}\,A_{i}^{\beta}({\bf r})\,
{\textstyle \frac{\partial_{i}}{\partial^{2}}}\left(\,
{\cal M}_{(\eta)}^{\vec{\alpha}}({\bf r})\,
\partial_{j}V_{j}^{\gamma}({\bf r})\,\right)
\nonumber \\
-\sum_{\eta=0}\sum_{t=1}&& (-1)^{t-1}g^{t+\eta}\,
{\textstyle \frac{B(\eta)}{\eta!(t-1)!(t+1)}}\,
f^{\vec{\mu}a\lambda}_{(t)}f^{\vec{\alpha}\lambda\gamma}_{(\eta)}\,
{\cal R}_{(t)}^{\vec{\mu}}({\bf r})\,
{\cal M}_{(\eta)}^{\vec{\alpha}}({\bf r})\,
\partial_{i}V_{i}^{\gamma}({\bf r})
\nonumber \\
-gf^{a\beta d}A_{i}^{\beta}({\bf r})&&\!
\sum_{\eta=0}\sum_{t=1} (-1)^{t}g^{t+\eta}\,
{\textstyle\frac{B(\eta)}{\eta!(t+1)!}}
f^{\vec{\mu}d\lambda}_{(t)}
f^{\vec{\alpha}\lambda\gamma}_{(\eta)}
{\textstyle\frac{\partial_{i}}{\partial^{2}}}\,
\left(\,{\cal R}_{(t)}^{\vec{\mu}}({\bf r})\,
{\cal M}_{(\eta)}^{\vec{\alpha}}({\bf r})\,
\partial_{j}V_{j}^{\gamma}({\bf r})\,\right)\;.
\label{eq:Asum}
\end{eqnarray}
where  ${\cal{R}}^{\vec{\alpha}}_{(\eta)}({\bf{r}})$ is the product of 
functions of the longitudinal gauge field
\begin{equation}
{\cal{R}}^{\vec{\alpha}}_{(\eta)}({\bf{r}})=
\prod_{m=1}^\eta{\cal{X}}^{\alpha[m]}({\bf{r}})\;\;\;\mbox{with}\;\;\; 
{\cal{X}}^\alpha({\bf{r}}) =
[\,{\textstyle\frac{\partial_i}{\partial^2}}A_i^\alpha({\bf{r}})\,]\;.
\label{eq:XproductN}
\end{equation}
Similarly, ${\cal{M}}_{(\eta)}^{\vec{\alpha}}({\bf{r}})$ is a functional of the resolvent gauge field
\begin{equation}
{\cal{M}}_{(\eta)}^{\vec{\alpha}}({\bf{r}})
=\prod_{m=1}^\eta
\overline{{\cal Y}^{\alpha[m]}}({\bf{r}})\;\;\;\mbox{with}\;\;\;
\overline{{\cal Y}^{\alpha}}({\bf r})=
{\textstyle \frac{\partial_{j}}{\partial^{2}}
\overline{{\cal A}_{j}^{\alpha}}({\bf r})}\,.
\label{eq:defM}
\end{equation}
Eq.(\ref{eq:Asum}) also contains the chain of SU(3) structure constants 
\begin{equation}
f^{\vec{\alpha}\beta\gamma}_{(\eta)}=f^{\alpha[1]\beta b[1]}\,
\,f^{b[1]\alpha[2]b[2]}\,f^{b[2]\alpha[3]b[3]}\,\cdots\,
\,f^{b[\eta-2]\alpha[\eta-1]b[\eta-1]}f^{b[\eta-
1]\alpha[\eta]\gamma}\;,
\label{eq:fproductN}
\end{equation}
the Bernoulli numbers $B({\eta})$, and $V_{j}^{\gamma}({\bf r})$, which 
represents an arbitrary vector field in the adjoint representation of SU(3).
As shown in our earlier work,~\cite{CBH2} Eq.(\ref{eq:Asum}) is a reformulation of the non-Abelian 
Gauss's law as a requirement on the resolvent gauge field $\overline {{\cal A}_{i}^{\gamma}}({\bf r})$. 
In order to implement Gauss's law, it is also necessary to find an explicit solution of 
Eq.(\ref{eq:Asum}). We provided such a solution in our work,~\cite{CBH2} but, for lack of space, cannot 
repeat it here. 
\section{Construction of the Gauge-Invariant Fields}
The gauge field, and the resolvent gauge field, play a crucial role in determining the gauge-invariant 
spinor (quark) and gauge (gluon) fields. As we pointed out in earlier work,~\cite{CBH2} the 
Gauss's law operator ${\hat {\cal G}}^{a}({\bf r})$, given in Eq.(\ref{eq:QCDGauss}), and 
the `pure glue' Gauss's law operator 
${\cal G}^{a}({\bf r})=\partial_{i}\Pi_{i}^{a}({\bf r})+gf^{abc}A_{i}^{b}({\bf r})
\Pi_{i}^{c}({\bf r})$, are 
unitarily equivalent, so that ${\cal G}^a$ may be taken to represent 
${\hat {\cal G}}^{a}({\bf r})$ in a different, 
unitarily equivalent representation. We refer to the representation in which 
${\hat {\cal G}}^{a}({\bf r})$ is the 
Gauss's law operator, the ${\cal C}$ representation; and the representation in which 
${\cal G}$ represents the {\em entire} Gauss's law 
operator, with the color-charge density $j^a_{0}({\bf{r}})$ included (though implicitly only),
the ${\cal N}$ representation. The unitary equivalence is expressed as~\cite{CBH2} 
\begin{equation}
{\hat {\cal G}}^{a}({\bf r})
={\cal{U}}_{\cal{C}}\,
{\cal G}^{a}({\bf r})\,{\cal{U}}^{-1}_{\cal{C}}\,,
\label{eq:Gtrans}
\end{equation} 
where ${\cal{U}}_{\cal{C}}$ is a functional of the gauge field and the resolvent gauge field, as well as 
the color-charge density: ${\cal{U}}_{\cal{C}}=e^{{\cal C}_{0}}
e^{\bar {\cal C}}$ with
${{\cal C}_{0}}$ and ${\bar {\cal C}}$ given by
\begin{equation}
{\cal C}_{0}=i\,\int d{\bf{r}}\,
{\textstyle {\cal X}^{\alpha}}({\bf r})\,j_{0}^{\alpha}({\bf r})\;,
\;\;\;\;\;\mbox{and}\;\;\;\;\;\;\;
{\bar {\cal C}}=i\,\int d{\bf{r}}\,
\overline{{\cal Y}^{\alpha}}({\bf r})\,j_{0}^{\alpha}({\bf r})\;.
\label{eq:CCbar}
\end{equation}
Since the quark field $\psi$ trivially commutes with ${\cal G}^{a}({\bf r})$, 
 $\psi$ is manifestly gauge-invariant in the 
${\cal N}$ representation. The unitary operator ${\cal{U}}_{\cal{C}}$ transforms the quark 
field $\psi$ so that the gauge-invariant spinor field in the ${\cal C}$ representation is
\begin{equation}
{\psi}_{\sf GI}({\bf{r}})={\cal{U}}_{\cal C}\,\psi({\bf{r}})\,{\cal{U}}^{-1}_{\cal C}=
V_{\cal{C}}({\bf{r}}){\psi}({\bf{r}})
\label{eq:psiGI}
\end{equation}
where
\begin{equation}
V_{\cal{C}}({\bf{r}})=
\exp\left(\,-ig{\overline{{\cal{Y}}^\alpha}}({\bf{r}})
{\textstyle\frac{\lambda^\alpha}{2}}\,\right)\,
\exp\left(-ig{\cal X}^\alpha({\bf{r}})
{\textstyle\frac{\lambda^\alpha}{2}}\right)\;.
\label{eq:el1}
\end{equation}
and ${\lambda}^h$ represents the Gell-Mann SU(3) matrices. And, since $V_{\cal{C}}({\bf{r}})$ is 
a unitary operator, we can also readily see that the corresponding gauge-invariant gauge (gluon) field 
$A_{{\sf GI}\,i}^{b}({\bf{r}})$ is given by
\begin{equation}
[\,A_{{\sf GI}\,i}^{b}({\bf{r}})\,{\textstyle\frac{\lambda^b}{2}}\,]
=V_{\cal{C}}({\bf{r}})\,[\,A_{i}^b({\bf{r}})\,
{\textstyle\frac{\lambda^b}{2}}\,]\,
V_{\cal{C}}^{-1}({\bf{r}})
+{\textstyle\frac{i}{g}}\,V_{\cal{C}}({\bf{r}})\,
\partial_{i}V_{\cal{C}}^{-1}({\bf{r}})\;,
\label{eq:gigf}
\end{equation}
or, equivalently,
\begin{equation}
A_{{\sf GI}\,i}^b({\bf{r}})=
A_{T\,i}^b({\bf{r}}) +
[\delta_{ij}-{\textstyle\frac{\partial_{i}\partial_j}
{\partial^2}}]\overline{{\cal{A}}^b_{j}}({\bf{r}})\,.
\label{eq:Adressedthree1b}
\end{equation}
We can expand ${\psi}_{\sf GI}({\bf{r}})$ and $A_{{\sf GI}\,i}^b({\bf{r}})$ 
to arbitrary order in $g$, and, in that way, we have verified that our gauge-invariant 
quark and gluon fields agree with the perturbative calculations of Lavelle, McMullan, $et.\,al.$ to 
the highest order to which their perturbative calculations were available.~\cite{lavelle2,lavelle3}  
\section{The Gauge-Invariant QCD Hamiltonian}
We can construct the gauge-invariant QCD Hamiltonian by systematically transforming the temporal-gauge
QCD Hamiltonian from the conventional ${\cal C}$ representation to the ${\cal N}$ representation.~\cite{BCH3}
In that representation, $\psi({\bf{r}})$ represents the {\em gauge-invariant} quark field, and 
$j^a_0({\bf{r}})$ the {\em gauge-invariant} color-charge density. In the ${\cal N}$ representation, 
$j^a_0({\bf{r}})$ therefore implicitly includes `glue-color' as well as the color of the bare quarks. 
\smallskip

The QCD Hamiltonian that results from the transformation to the ${\cal N}$ representation is 
\begin{equation}
\tilde{H}=\int d{\bf r}\left[ \ {\textstyle \frac{1}{2}}
\Pi^{a}_{i}({\bf r})\Pi^{a}_{i}({\bf r})
+  {\textstyle \frac{1}{4}} F_{ij}^{a}({\bf r}) F_{ij}^{a}({\bf r})+
{\psi^\dagger}({\bf r})\left(\beta m-i\alpha_{i}
\partial_{i}\right)\psi({\bf r})\right] + \tilde{H}^{\prime}\,.
\label{eq:HQCDN}
\end{equation}
$\tilde{H}^{\prime}$ describes interactions involving the gauge-invariant quark field. The 
parts of $\tilde{H}^{\prime}$ relevant to the dynamics of quarks and gluons can be expressed as 
\begin{equation}
\tilde{H}^{\prime}=\tilde{H}_{j-A}+\tilde{H}_{LR}\,.
\label{eq:Hprime}
\end{equation} 
$\tilde{H}_{j-A}$ describes the interaction of the gauge-invariant gauge field with the 
gauge-invariant quark color-current density, and is given by 
\begin{equation}
\tilde{H}_{j-A}=-\,g\int d{\bf r}\,{\psi^\dagger}({\bf r})\alpha_{i}{\textstyle\frac{\lambda^h}{2}}\psi({\bf r})\,
A_{{\sf GI}\,i}^{h}({\bf r})\,;
\label{eq:HJA}
\end{equation}
As in the case of QED, $\tilde{H}_{j-A}$ couples the gauge-invariant gauge field to the current density
$j^h_i({\bf r})=g{\psi^\dagger}({\bf r})\alpha_{i}{\textstyle\frac{\lambda^h}{2}}\psi({\bf r})$, and can 
be expected to have a $v/c$ dependence for non-relativistic quarks, leaving the nonlocal 
${\tilde{H}}_{LR}$ as the dominant term in the low-energy regime.
${\tilde{H}}_{LR}$ is the nonlocal interaction 
\begin{equation}
{\tilde{H}}_{LR}=H_{g-Q}+H_{Q-Q}\,.
\label{eq:HLR}
\end{equation}
 A useful 
formulation of $H_{Q-Q}$ can be given as~\cite{CH1}
\begin{equation}
H_{Q-Q}=\frac{1}{2}{\int}d{\bf r}d{\bf x}\,{j}_0^b({\bf r}){\cal F}^{ba}({\bf r},{\bf x}){j}_0^a({\bf x})
\label{eq:HQQalt}.
\end{equation} 
where the Green's function ${\cal F}^{ba}({\bf r},{\bf x})$ is represented as 
\begin{eqnarray} 
{\cal F}^{ba}({\bf r},{\bf x})=\frac{{\delta}_{ab}}{4{\pi}|{\bf r}-{\bf x}|}
&&\!\!+\,2g\,f^{{\delta}_{(1)}ba}\int\frac{d{\bf y}}{4{\pi}|{\bf r}-{\bf y}|}{A_{{\sf GI}\,i}^
{{\delta}_{(1)}}({\bf{y}})\,\partial_{i}\frac{1}{4{\pi}|{\bf y}-{\bf x}|}}+\nonumber \\
{\cdots}+(-1)^{(n-1)}(n+1)g^n&&\!\!\!f^{{\delta}_{(1)}bs_{(1)}}{\cdots}f^{s_{(n-1)}{\delta}_{(n)}a}
\int\frac{d{\bf y}_1}{4{\pi}|{\bf r}-{\bf y}_1|}
{A}_{{\sf GI}\,i}^{{\delta}_{(1)}}({\bf{y}_1})\,\nonumber \\
\partial_{i}\int\!\!\frac{d{\bf y}_2}{4{\pi}|{\bf y}_1-{\bf y}_2|}\,
\cdots&&\int\frac{d{\bf y}_n}{4{\pi}|{\bf y}_{(n-1)}-{\bf y}_n|}\,
{A}_{{\sf GI}\,{\ell}}^{{\delta}_{(n)}}
({\bf y}_{n})\,\partial_{\ell}
\frac{1}{4{\pi}|{\bf y}_n-{\bf x}|}\,.
\label{eq:Lexp}
\end{eqnarray}
We make the following observations about ${\cal F}^{ba}({\bf r},{\bf x})$: 
The initial term resembles the Coulomb interaction. The infinite series of further terms consists of chains 
through which the interaction is transmitted from one 
color-charge density to the other. Each
chain contains a succession of `links', which have the characteristic form
\be
\mbox{link}=gf^{s_{(1)}{\delta}s_{(2)}}
{A}_{{\sf GI}\,j}^{{\delta}}
({\bf x})\,\partial_{j}\frac{1}{4{\pi}|{\bf x}-{\bf y}|}\,.
\label{eq:link}
\ee 
The `links' are coupled through summations over the $s_{(n)}$ indices and 
integrations over the spatial variables. All the quantities in $H_{Q-Q}$ are gauge-invariant 
--- the color-charge density 
${j}_0^b({\bf r})$ as well as the gauge field ${A}_{{\sf GI}\,j}^{{\delta}}({\bf x})$.
$H_{g-Q}$ --- the other nonlocal interaction in ${\tilde{H}}_{LR}$ ---
couples quark to gauge-invariant gluon color. 
In $H_{g-Q}$ the quark color-charge density is coupled, through the same Green's function 
${\cal F}^{ba}({\bf r},{\bf x})$, to a gauge-invariant expression describing `glue'-color, 
given by  ${\sf K}_g^d({\bf r})=gf^{d\sigma e}\,{\sf Tr}\left[V_{\cal{C}}^{-1}({\bf{r}})
{\textstyle\frac{\lambda^e}{2}}V_{\cal{C}}({\bf{r}}){\textstyle\frac{\lambda^b}{2}}\right]
A_{{\sf GI}\,i}^{\sigma}({\bf r})
\Pi^{b}_{i}({\bf r})\,.$ \smallskip

A number of further observations about Eq.(\ref{eq:Lexp}) have relevance for low-energy QCD dynamics. The first of 
these is based on a kind of `color multipole' expansion of 
${\int}{\cal F}^{ba}({\bf r},{\bf x}){j}_0^a({\bf x})d{\bf x}$ --- which appears 
as part of $H_{Q-Q}$ as well as $H_{g-Q}$ --- about the point ${\bf x}={\bf x}_0$:
\begin{equation} 
{\int}d{\bf x}\,\left\{{\cal F}^{ba}({\bf r},{\bf x}_0)+X_i\partial_i
{\cal F}^{ba}({\bf r},{\bf x}_0)+\frac{1}{2}\,X_iX_j\,\partial_i\partial_j
{\cal F}^{ba}({\bf r},{\bf x}_0)\;+\;\;\cdots\right\}{j}_0^a({\bf x})
\label{eq:Lmultipole}
\end{equation}
where $X_i=(x-x_0)_i$ and $\partial_i=\partial /\partial x_i\,.$ When we perform the integration in 
Eq.~(\ref{eq:Lmultipole}), the first term contributes ${\cal F}^{ba}({\bf r},{\bf x}_0)\,
{\cal Q}^a\,,$ where 
${\cal Q}^a=\int\,d{\bf x}\,{j}_0^a({\bf x})$ (the integrated `color charge'). 
Since the color charge is the 
generator of infinitesimal rotations in SU(3) space, 
it will annihilate any multiquark state vector in a singlet color configuration. 
Multiquark packets in a singlet color configuration therefore are immune to 
the initial term of the nonlocal 
$H_{Q-Q}$. Color-singlet configurations of quarks are only subject to the color multipole terms, 
which act as color analogs to the Van der Waals interaction.
The scenario that this model suggests is that the leading term in $H_{Q-Q}$, namely 
${\cal Q}^b{\cal F}^{ba}({\bf r}_0,{\bf x}_0){\cal Q}^a$ for a quark color charge ${\cal Q}^a$ at ${\bf r}_0$ and 
another quark color charge ${\cal Q}^b$ at ${\bf x}_0$, contributes to the confinement of 
quarks and packets of quarks that are not in color-singlet configurations. 
Moreover, assuming that ${\cal F}^{ba}({\bf r},{\bf x}_0)$ 
varies only gradually within a volume occupied by quark packets, the effect of the higher order
color multipole forces on a packet of quarks in a color-singlet configuration becomes more significant as 
the packet increases in size. As small quark packets move through gluonic matter, they 
will experience only insignificant effects from the multipole contributions to $H_{Q-Q}$, 
since, as can be seen from 
Eq.~(\ref{eq:Lmultipole}), the factors $X_i\,$, $X_iX_j$, ${\cdots}$, $X_{i(1)}{\cdots}X_{i(n)}$,
keep the higher order multipole terms from making significant contributions to 
${\int}d{\bf x}\,{\cal F}^{ba}({\bf r},{\bf x}){j}_0^a({\bf x})$ when they are integrated 
over small packets of quarks. 
As the size of the quark packets increases, the regions
over which the multipoles are integrated also increases, and the effect of the multipole interactions
on the color-singlet packets can become larger. 
This dependence on packet size of the final-state interactions experienced by color-singlet states ---
$i.\,e.$ the increasing importance of final-state interactions as color-singlet packets grow in size --- 
is in qualitative agreement with the characterizations of color transparency and color coherence 
given by Miller and by Jain, Pire and Ralston.~\cite{miller,ralston} \smallskip

Another observation pertains to the series given in Eq.(\ref{eq:Lexp}). To fully achieve the purpose of this 
investigation, it is necessary to evaluate ${\cal F}^{ba}({\bf r},{\bf x})$ nonperturbatively. 
The $n^{th}$ order term of ${\cal F}^{ba}({\bf r},{\bf x})$ --- $i.\,e.$ the chain with $n$ links ---
has such a regular structure, that an explicit 
form can easily be written, without requiring knowledge of the lower order terms in the series. 
A nonperturbative evaluation of  ${\cal F}^{ba}({\bf r},{\bf x})$ is therefore far more accessible here than 
in the `standard' gauge-dependent formulation in the ${\cal C}$ representation, in which it is much more
difficult to identify and isolate the interactions that make the dominant contributions in the low-energy
regime. \smallskip

In order to carry out a nonperturbative evaluation of $H_{Q-Q}$, either the ${\cal F}^{ba}({\bf r},{\bf x})$
series has to be formally summed, or a more indirect method has to be found to achieve that objective.
As of now, we have not yet addressed the evaluation of the 
gauge-invariant gauge field, which is an important component of ${\cal F}^{ba}({\bf r},{\bf x})$.
We have investigated an SU(2) model 
of QCD under the assumption that gluon correlations can be neglected, so that the expectation
value ${\langle}0|A_{{\sf GI}\,i}^{a}({\bf r})|0\rangle$, in a state that implements the Gauss's law 
${\cal G}^a({\bf r})|0\rangle=0$, can be substituted for the operator-valued 
$A_{{\sf GI}\,i}^{a}({\bf r})$ in Eq.(\ref{eq:Lexp}).  Furthermore, we made the {\em ansatz} 
that the transverse
${\langle}0|A_{{\sf GI}\,i}^{a}({\bf r})|0\rangle$ can be modeled by the manifestly transverse SU(2) 
`hedgehog' function 
${\langle}A_{{\sf GI}\,i}^{\delta}({\bf{r}})\rangle={\epsilon}^{ij{\delta}}r_j{\phi}(r).$
In this work, we showed how the SU(2) algebra enabled us to reduce the evaluation of 
${\cal F}^{ba}({\bf r},{\bf x})$ to the solution of a sixth-order differential equation,~\cite{CH1}
which circumvented the need for a perturbative evaluation of ${\cal F}^{ba}({\bf r},{\bf x})$.

\section{Discussion}
Abelian and non-Abelian gauge theories resemble each other in an important respect: The common 
thread that connects them is that the use of gauge-invariant fields eliminates interactions of pure-gauge 
components of gauge fields with charge and current densities, and replaces those interactions with nonlocal  
interactions among charge densities --- the Coulomb interaction in QED, and ${\tilde H}_{LR}$,
given in Eq.(\ref{eq:HLR}), in QCD.
The structure of the Green's function ${\cal F}^{ba}({\bf r},{\bf x})$ --- a series of chains consisting of 
coupled links, in which the $n^{th}$ order chain is a product of $n$ links --- 
suggests features closely associated with QCD: flux tubes, `string'-like structures tying 
colored objects to each other, etc.  Further work will be required to determine whether these suggestive 
analogies are supported by detailed calculations.

\section*{Acknowledgments}
This research was supported by the Department of Energy
under Grant No. DE-FG02-92ER40716.00.

\end{document}